\documentclass[aps,prd,tightenlines,preprintnumbers,showpacs,superscriptaddress,nofootinbib,eqsecnum,floatfix,amsmath,longbibliography,
twocolumn]{revtex4-2}
\usepackage{bm}
\usepackage{ulem}
\usepackage{graphicx}
\usepackage{comment}
\usepackage[pdftex]{color}

\usepackage[dvipsnames]{xcolor}

\definecolor{dblue}{rgb}{0.0, 0.0, 0.6}
\definecolor{ddblue}{rgb}{0.0, 0.0, 0.4}
\definecolor{dgreen}{rgb}{0.0, 0.4, 0.0}
\definecolor{dgray}{gray}{0.4}
\definecolor{ddgray}{gray}{0.3}

\usepackage{hyperref}

\def\chib{{\overline{\chi}}}

\def\Fig#1{Fig.~(\ref{#1})}

\newcommand{\KD}{{K\"{a}hler-Dirac }}

\definecolor{orange}{rgb}{1.0, 0.5, 0}

\begin{document}
\preprint{FERMILAB-PUB-24-0550-V}
\title{Symmetric Mass Generation with four SU(2) doublet fermions}
\author{Nouman Butt}
\affiliation{Department of Physics,  University of Illinois, Urbana Champagne USA}
\author{Simon Catterall~\thanks{smcatter@syr.edu}}
\affiliation{Department of Physics, Syracuse University NY 13244 USA}
\author{Anna Hasenfratz~\thanks{Email: Anna.Hasenfratz@colorado.edu}}
\affiliation{Department of Physics, University of Colorado Boulder,  CO 80309, USA}

\begin{abstract}
We study a single exactly massless
staggered fermion in the fundamental representation of an $SU(2)$ gauge group. 
We utilize an nHYP-smeared fermion action supplemented with additional heavy Pauli-Villars fields
which serve to decrease lattice artifacts. The phase diagram exhibits a clear two-phase structure
with a conformal phase at weak coupling and a novel new phase, the Symmetric Mass Generation (SMG) phase, appearing at strong coupling.
The SMG phase is confining with all states gapped and chiral symmetry
unbroken. Our finite size scaling analysis provides
strong  evidence that the phase transition between these two phases is continuous, which would
allow for the existence of a continuum SMG phase. Furthermore, the RG flows are consistent
with a $\beta$-function that vanishes quadratically at the new fixed point suggesting that the $N_f=4$ flavor SU(2) gauge theory lies at the opening of the conformal window.
\end{abstract}

\maketitle

\newpage

\section{Introduction}\label{sec:motivation}

Symmetric Mass Generation (SMG) is a conjectured non-perturbative mechanism for giving mass
to fermions without breaking chiral symmetries. The mass arises not from
a pairing of elementary fermions as it would arise in a classical Lagrangian but as a consequence
of the presence of a non-trivial vacuum state consisting of a symmetric multi-fermion condensate
\cite{Butt:2018nkn,Catterall:2017ogi,Butt:2021brl,Wang:2022ucy,Slagle:2014vma,Chandrasekharan:2013aya,Ayyar:2014eua,Ayyar:2015lrd,Ayyar:2016lxq}.
A necessary condition for a theory to possess
an SMG phase is
that all 't Hooft anomalies of the theory must vanish, since otherwise, a nonzero anomaly would necessitate spontaneous symmetry breaking in the IR
\cite{Garcia-Etxebarria:2018ajm,Razamat:2020kyf,Catterall:2022jky}.

To fully understand the constraints needed for SMG one must generalize the notion of symmetry
and anomaly to embrace both discrete symmetries and non-local symmetries that act on extended objects - so-called generalized symmetries \cite{Gaiotto:2014kfa,Brennan:2023mmt}. For example, in the presence of certain four fermion interactions, a four dimensional chiral theory may possess only a discrete spin-$Z_4$ global symmetry under which Weyl fields flip sign according to their chirality. 
The associated Dai-Freed anomaly can be computed and is canceled only if the theory contains
multiples of sixteen chiral fermions \cite{Garcia-Etxebarria:2018ajm, Razamat:2020kyf}.
This result agrees with arguments in condensed matter physics concerning the number of fermions
needed to gap out edge states in topological superconductors \cite{Fidkowski:2009dba,You:2014oaa}. It also agrees with the cancellation of gravitational 't Hooft anomalies for K\"{a}hler-Dirac fermions \cite{Catterall:2023nww,Catterall:2022jky}.

\KD fermions are of particular interest, since, when discretized on a regular torus, they yield staggered fermions.  Furthermore, the mixed $Z_4$-gravitational anomaly of the continuum \KD theory
survives intact under discretization yielding an exact constraint on the number of staggered
fermions needed for an SMG phase. The focus of this paper is a search for such an SMG phase
in the minimal strongly coupled staggered gauge theory with vanishing
$Z_4$ 't Hooft anomaly -- a single massless 
staggered fermion coupled to an $SU(2)$ gauge field. In the continuum limit and deep in
the ultraviolet (UV) this lattice theory
corresponds to exactly eight Dirac fermions or sixteen Majorana fermions.

Prior work by Hasenfratz \cite{Hasenfratz:2022qan} suggests that the SU(3) gauge theory with 8 massless Dirac fermions, represented by two sets of staggered flavors, indeed exhibits such an SMG phase at strong gauge couplings. In weak coupling the system appears conformal, and the transition between the two phases appears to be continuous.
Our investigations of the SU(2) gauge system with a single staggered fermion show very similar properties. In a high statistics numerical simulation we establish that the system is conformal in the weak coupling regime but gapped yet chirally symmetric at strong coupling. The phase transition separating them appears continuous in a finite size scaling analysis. Even more striking is the observed renormalization group (RG) flow 
which suggests that the phase transition is governed by a FP that is IR attractive at weak coupling and IR repulsive at strong coupling. This rather unusual scenario can be realized when two fixed points merge (mFP) \cite{Kaplan:2010zz,Vecchi:2010jz,Gorbenko:2018ncu}. 

\begin{figure}[tbh]
    \centering
    \includegraphics[width=0.235\textwidth]{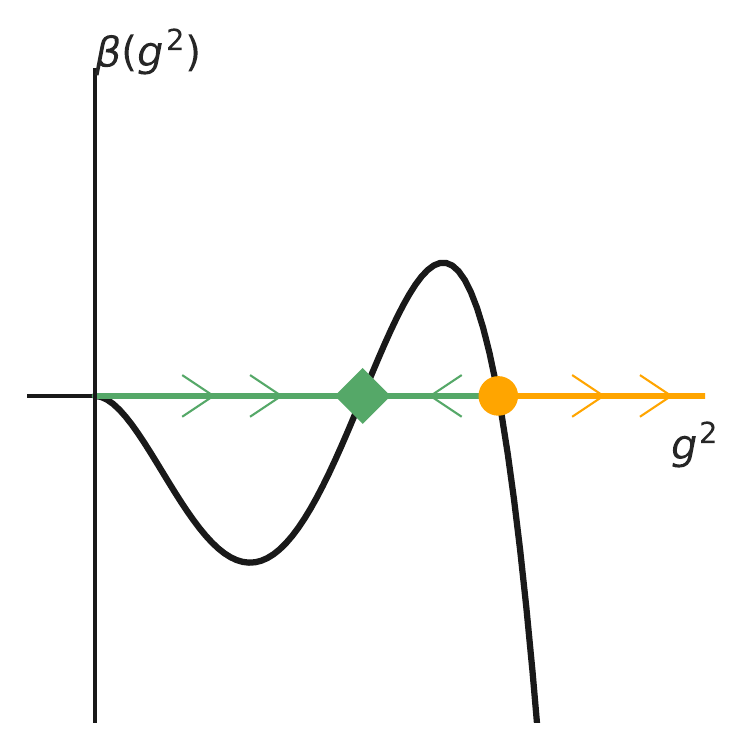}
    \includegraphics[width=0.235\textwidth]{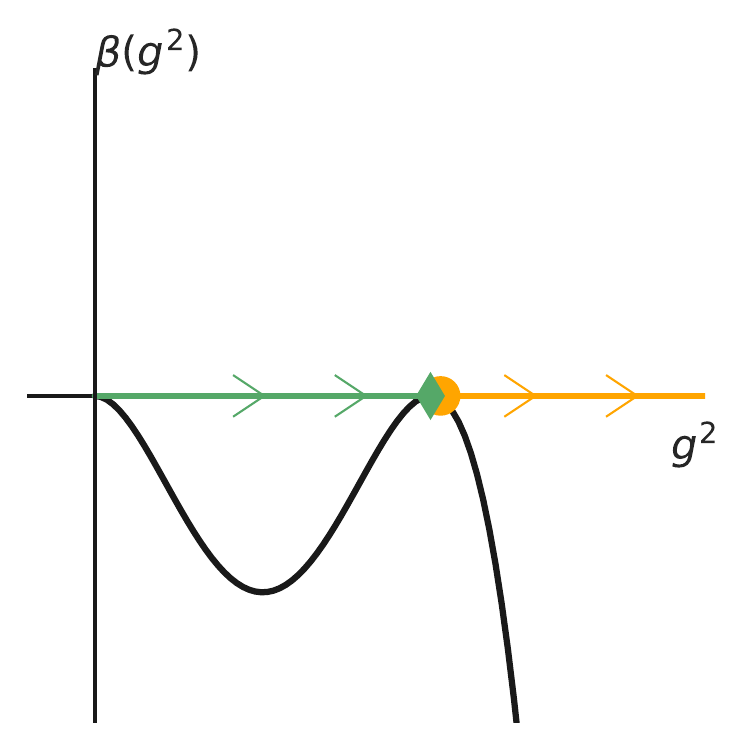}\\
    \caption{Sketches of possible RG $\beta$ functions. The left panel depicts a theory
    with an IR conformal phase at weak coupling (green) together with a strongly coupled phase where the gauge coupling is relevant (orange). The right panel shows a possible opening of the conformal window via merged fixed points (mFP).}
    \label{fig:bfn-sketch}
\end{figure}

In \Fig{fig:bfn-sketch} we sketch two possibilities for the RG $\beta$ function. The left panel refers to a system inside the conformal window with an infrared attractive fixed point (IRFP)  denoted by a green diamond. Anywhere within the weak coupling regime (indicated by the green line) the system flows into the IRFP. The orange circle signals the presence of an ultraviolet attractive fixed point (UVFP), separating the conformal phase from a strongly coupled phase - the latter is the conjectured SMG phase in our system.
The right panel of \Fig{fig:bfn-sketch} exhibits a special case where the two
fixed points merge into a single fixed point. Such a scenario may arise at the opening
of the conformal window as suggested in \cite{Kaplan:2010zz,Vecchi:2010jz,Gorbenko:2018ncu}. The new FP is IR attractive at weak coupling but UV attractive (IR unstable)
at strong coupling. We refer to this possibility as the merged FP (mFP) scenario.
In both panels, the arrows indicate the direction of the RG flow from UV to IR.

The numerical results we present in the rest of this paper favor the mFP scenario corresponding to the right panel of \Fig{fig:bfn-sketch}. However, we cannot exclude the possibility of two separate FPs very close to each other. If the $N_f=4$ flavor SU(2) gauge theory indeed resides precisely
at the opening of the conformal window, it suggests that the theory is rather special. We conjecture
that this feature is related to the fact that it is
the gauge theory with the minimal number of flavors needed to achieve a vanishing $Z_4$ 't Hooft
anomaly. 

\section{Model}

Our numerical setup is nearly identical to the one used in Refs. \cite{Hasenfratz:2022qan,Hasenfratz:2024fad}. The gauge action is a combination of fundamental and adjoint plaquettes with relative coefficients $\beta_A/\beta_F=-0.25$ \cite{Cheng:2011ic}. In this paper we refer to the bare coupling of the fundamental plaquette as $\beta_b (\propto 4/g^2_0)$  to distinguish it from the RG $\beta$ function.
The action takes the form
\begin{align}
    S&=\sum_{x,\mu}\chib(x)\left(V_\mu(x)\chi(x+\mu)-V_\mu^\dagger(x-\mu)\chi(x-\mu)\right)\nonumber\\& + S_{\rm gauge}+S_{PV}
\end{align}
where the gauge links $V_\mu(x)$ are nHYP smeared with smearing coefficients ($0.5,0.5,0.4$)~\cite{Hasenfratz:2002jn,Hasenfratz:2007rf}, and the fermions are SU(2) doublets.
In addition to translation, rotation and SU(2) gauge invariance the action
is invariant under a $U(1)$ symmetry in which the staggered fermions transform as
\begin{align}
    \chi(x)\to e^{i\alpha\epsilon(x)}\chi(x)\, , \quad  
    \chib(x)\to e^{i\alpha\epsilon(x)}\chib(x)
\end{align}
where $\epsilon(x)$ is the site parity $\epsilon(x)=\left(-1\right)^{\sum_i x_i}$. If the
theory is formulated on a discretization of the sphere this symmetry is broken to $Z_4$. In \cite{Catterall:2022jky} it is argued that this $Z_4$
suffers from a mod 2 't Hooft anomaly which can be cancelled if the theory
consists of multiples of two staggered fields. 

Most many-flavor systems within the conformal window show first order bulk phase transitions induced by large vacuum fluctuations. This prevents simulations from reaching strong enough coupling where possible ultraviolet fixed points (UVFP) might control the dynamics. These unphysical bulk phase transitions can be removed with improved gauge actions. 
In our simulations we include eight sets of SU(2) doublet Pauli-Villars (PV) staggered bosons with mass $a m_{PV}=0.75$ to achieve this \cite{Hasenfratz:2021zsl}. The heavy PV bosons decouple in the infrared, their only role is to reduce cutoff effects. 

In this work, we map out the phase structure of this system by scanning the bare parameter space in a wide coupling range. We performed simulations at 20-25 $\beta_b$ coupling values on 
$L^2\times(2L)$ volumes, $L/a=16$, 24,32, and 36.  We collected up to 500 thermalized configurations on the smaller volumes separated by 10 molecular dynamics time units (MDTU), while on the larger volumes, we have  150-300 thermalized configurations.

We have calculated the correlators of several meson states: two that couple to the pseudoscalar channels (PS,PS2) and an additional one that couples to the vector (V), together
with their parity partners (S,S2,A). The explicit forms of the corresponding correlators and their usual Dirac structure are given in Table (1) 
in the Supplementary Materials.

\section{The phase diagram}

\subsection{The meson spectrum \label{sect:spectrum}}
\begin{figure}[tbh]
    \centering
    \includegraphics[width=0.495\textwidth]{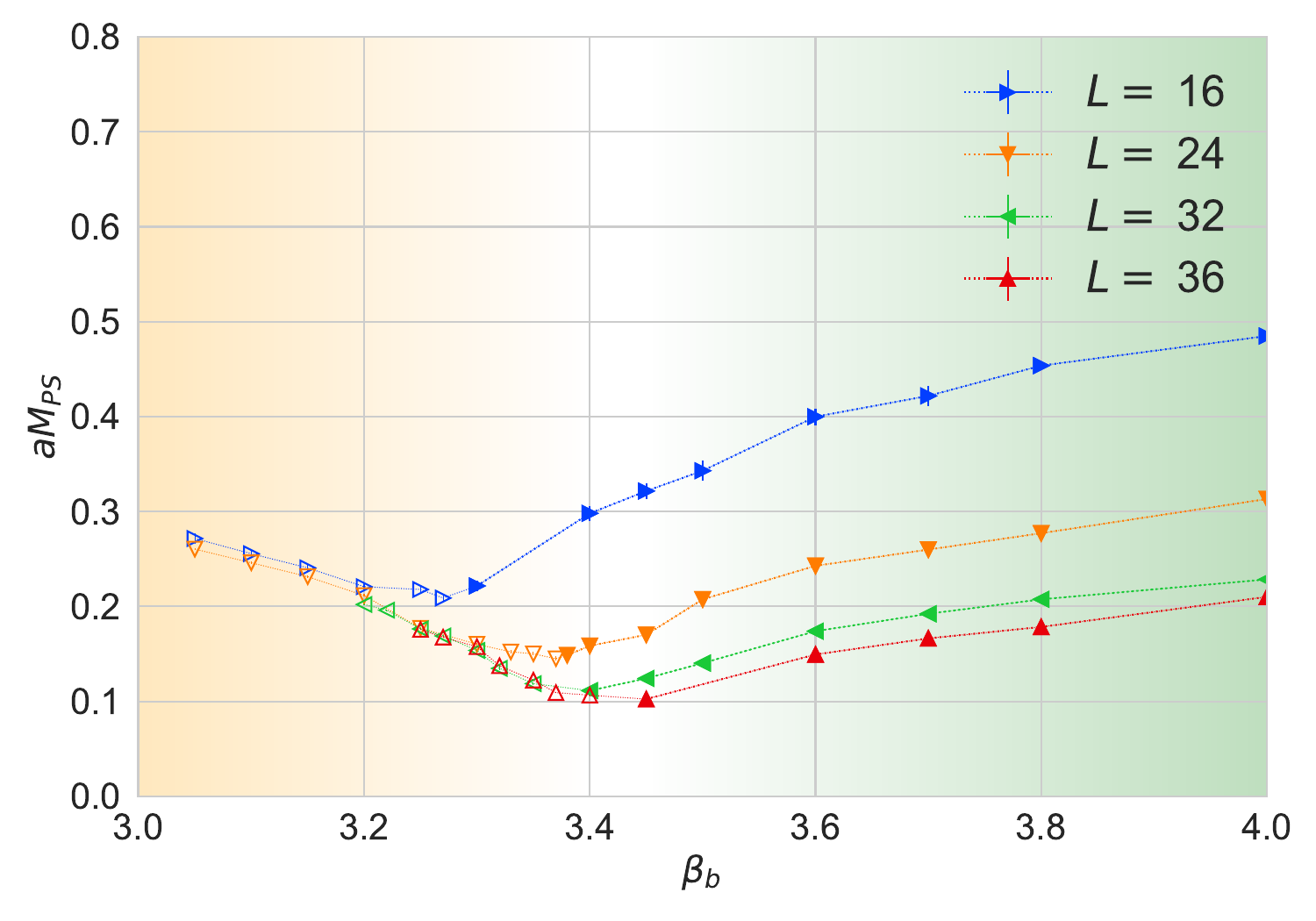}\\
    \caption{$a\,M_{PS}$, the mass of the would-be Goldstone boson in lattice units, as the function of the bare gauge coupling $\beta_b$ on volumes $L/a=16$, 24, 32, and 36. }
    \label{fig:Spectrum}
\end{figure}

\begin{figure*}[tbh]
    \centering
    \includegraphics[width=0.495\textwidth]{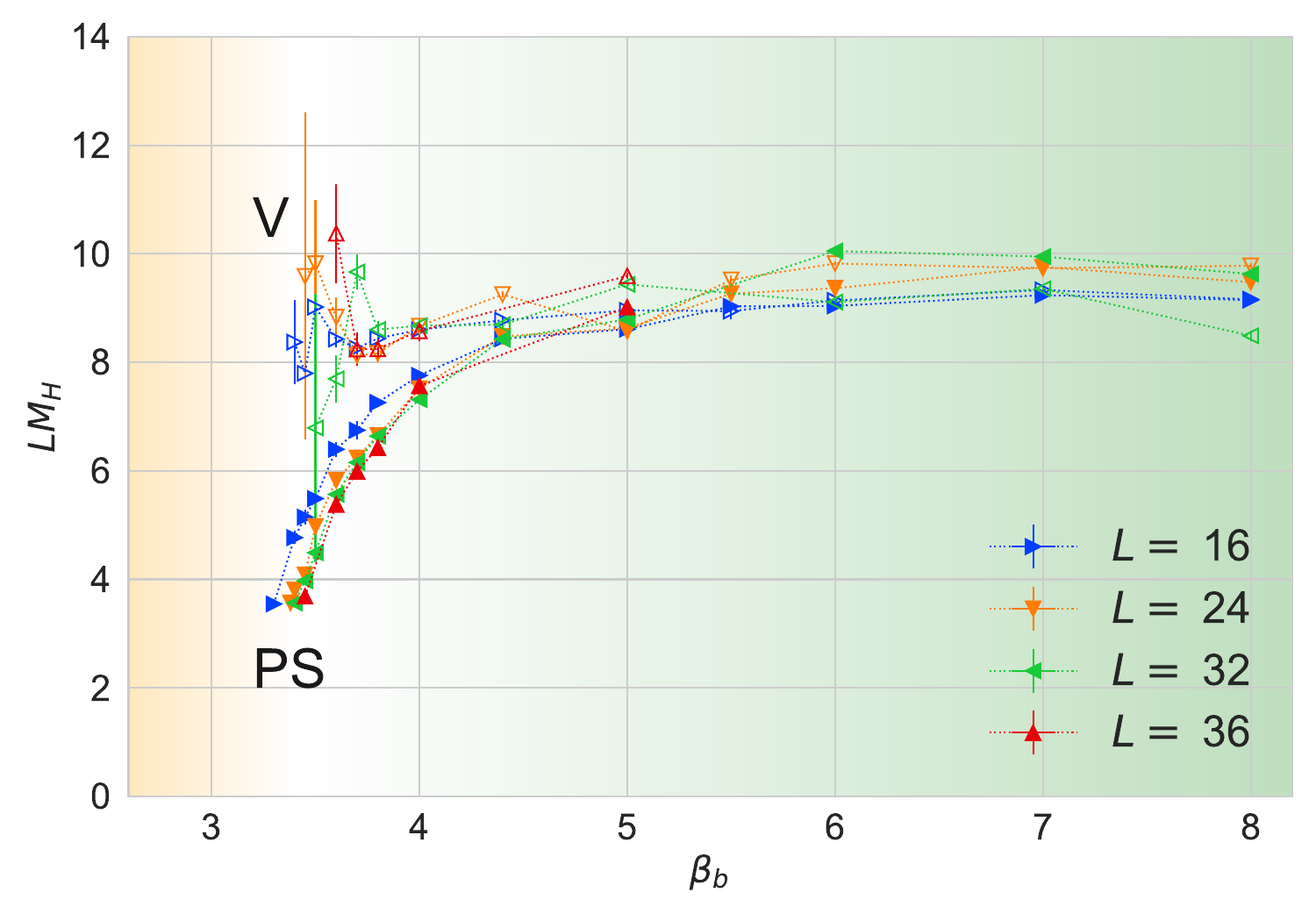}
    \includegraphics[width=0.495\textwidth]{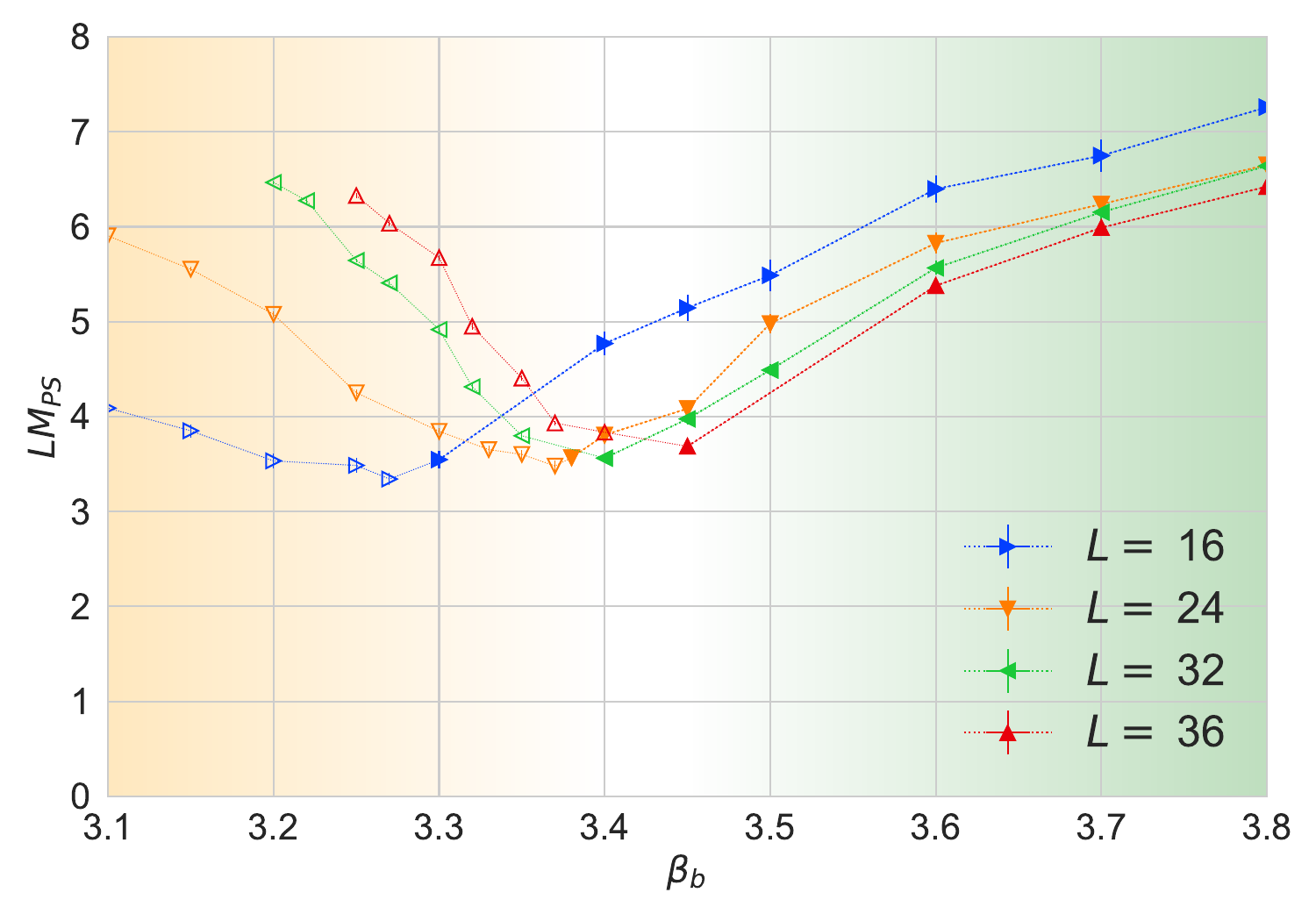}\\
    \caption{Left panel: $L\, M_{PS}$ and $L\, M_V$  as a function of the bare gauge coupling $\beta_b$ on volumes $L/a=16$, 24, 32, and 36. Right panel: $L\, M_{PS}$ vs $\beta_b$ zoomed into the critical region.}
    \label{fig:Vol_spectrum}
\end{figure*}

The volume dependence of the meson states can distinguish the different
phases of the system.  \Fig{fig:Spectrum} shows the mass of the PS meson (the pion) as a function of the bare coupling $\beta_b$ on several lattice volumes. 
In the strong coupling regime, indicated by yellow shading and open symbols, the mass is largely independent of the volume, implying a gapped phase. As the bare coupling $\beta_b$ increases, the smaller volumes peel off and undergo a finite-volume transition. 
This occurs when the lattice correlation length becomes comparable to the lattice size $L$. The infinite volume phase transition $\beta_b^*$ can be predicted by finite size scaling as we discuss in Sect.\ref{sect:FSS}. 
In contrast, within the weak coupling regime, indicated by green shading and filled symbols, we observe strong volume dependence consistent with conformal hyperscaling $M_{PS} \propto 1/L$. This scaling behavior holds not only for the lightest pseudoscalar state, but for all other mesons we consider. The left panel of \Fig{fig:Vol_spectrum} shows $L\,M_{PS}$ and also the vector state, $L\,M_{V}$, in the weak coupling regime. Both states are consistent with conformal hyperscaling, and nearly degenerate at large  $\beta_b$ (weak coupling). 
As the gauge coupling becomes stronger, a small volume dependence opens up. In the right panel of \Fig{fig:Vol_spectrum} we zoom into the critical region to show this for the PS state.

An RG transformation with scale factor $b>1$ induces a flow in the bare parameter
space and reduces both the dimensionless lattice correlation length and $L/a$ by a factor of $b$. The dimensionless quantity $L\,M_{PS}(\beta_b,L)$ is RG invariant;
this implies that $\beta_b$ must be adjusted as the volume is changed to keep $L\,M_{PS}$ fixed.
Indeed we can infer the flow in bare coupling from UV to IR by reading off the change of $\beta_b$ 
that is needed to keep $LM_{PS}$ fixed as the volume is reduced from large to small values.
For a conventional UVFP one expects to flow away from the critical region - that is what we observe in the SMG phase of \Fig{fig:Vol_spectrum}. 

The flow in the weak coupling phase  {\it does not} follow this trend. Smaller volumes appear to the left of the larger ones at fixed $LM_{PS}$; the direction of the RG flow is the same in the weak coupling and strong coupling regimes, suggesting that the critical point is IR attractive in the weak coupling side, corresponding to the mFP scenario.

The finite volume renormalized gradient flow coupling $g^2_{GF}$ \cite{Narayanan:2006rf,Luscher:2009eq,Luscher:2010iy} is another RG invariant quantity that can predict the direction of the RG flow. 
Fig.~(1) in the Supplementary Material shows  $g^2_{GF}(L;t)$ at gradient flow time $\sqrt{8t}= 0.5 \,L$ in the vicinity of the phase transition. While this quantity is monotonic in the bare coupling $\beta_b$, it shows the same RG flow as $L\,M_{PS}$: always to the left. This further supports the mFP scenario.

\subsection{The nature of the phase transition \label{sect:FSS}}
\begin{figure}[tbh]
    \centering
    \includegraphics[width=0.495\textwidth]{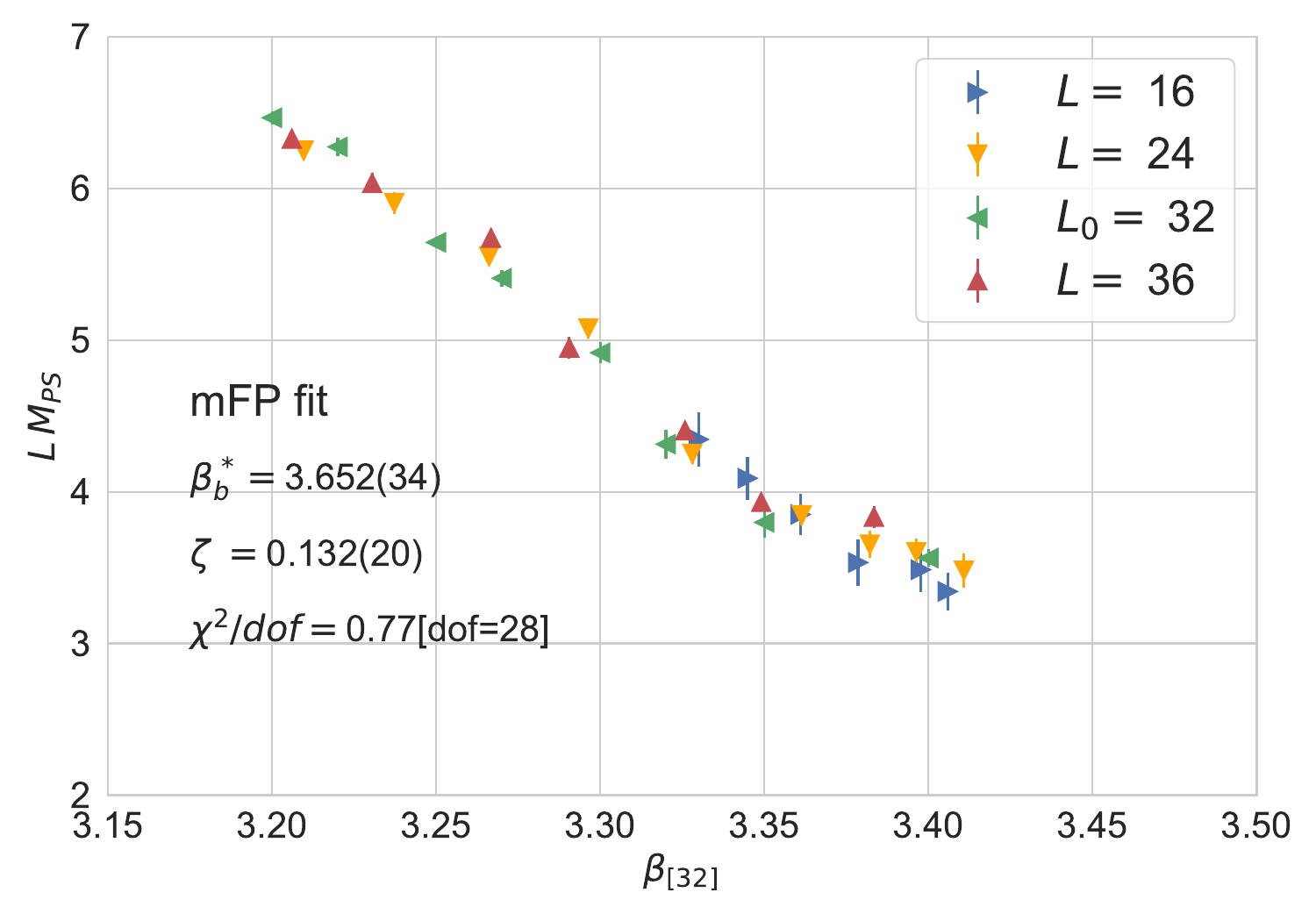}
    \caption{Finite size scaling plot using the operator $L\, M_{PS}$, assuming mFP scaling. We present the results in terms of the coupling $\beta_b$ of the $L_0=32$ reference volume according to Eq. \ref{eq:scaling2}. }
    \label{fig:FSS}
\end{figure}

In Sect. \ref{sect:spectrum} we have established that the SU(2) gauge theory coupled
to one massless staggered fermion exhibits two distinct phases. Furthermore, we argued
that one plausible scenario for understanding the phase transition is given by the mFP
scenario of Fig.~\ref{fig:bfn-sketch}. 
In this section, we investigate this possibility in more quantitative detail
employing standard finite-size scaling to the data. In this way we can attempt
to distinguish an mFP fixed point from a conventional second order phase transition (and exclude
the possibility of a first order transition).

We distinguish between
three possible scenarios: 
\begin{enumerate}
\item If the phase transition is second order, the correlation length scales as  
$\xi \propto |\beta_b /\beta^*_b -1|^{-\nu}$ where  $\beta^*_b$ denotes the critical coupling and $\nu$ is a universal critical exponent
\item If the phase transition is first order, the correlation length remains finite at the phase transition, but in finite volume the previous scaling formula is valid with $\nu=1/d=0.25$
\item If the phase transition corresponds to an mFP, described by a quadratic RG $\beta$ function $\beta_{R}(\beta_b) \propto (\beta_b-\beta^*_b)^2$ near criticality, the correlation length scales as 
$ \xi \propto e^{\zeta /|\beta_b /\beta^*_b - 1|}$ for some constant $\zeta$.
\end{enumerate}

At a fixed point with only one relevant direction, the correlation length is the only independent dimensional quantity. In a finite volume all scale dependence can then be expressed as a function of $\mathcal{X}=L/\xi$, i.e.
dimensionless operators exhibit unique scaling forms
\begin{equation}
\mathcal{O}(\beta_b,L) = F_\mathcal{O} (\mathcal{X})
 \label{eq:scaling}   
\end{equation}
where the function $F_\mathcal{O}$ depends on the operator. The scaling variables are
\begin{equation}
    \mathcal{X} =\begin{cases} L^{1/\nu} |\beta_b /\beta^*_b -1| \quad \text{for second/first order} \\
    L e^{\zeta /|\beta_b /\beta^*_b - 1|} \quad\quad\,\,\, \text{for mFP transition.}
    \end{cases}
    \label{eq:scaling2}
\end{equation}

Hasenfratz in Ref. \cite{Hasenfratz:2022qan} used the dimensionless finite volume gradient flow coupling \cite{Luscher:2010iy,Fodor:2012td} to investigate the critical behavior of the SU(3) gauge system with $N_f=8$ flavors. While a similar analysis is possible in our case, we choose to use a different quantity, the RG invariant combination $L\, M_{PS}(\beta_b;L)$. 

We perform the FSS analysis following the Nelder-Meade method \cite{Nelder:1965zz,Bhattacharjee:2001}. We choose a reference volume $L_0$ and use cubic spline interpolation to predict the reference curve $L_0\, M_{PS}(\beta_b;L_0)$ vs $\beta_b$.  Next we map the bare couplings $\beta_b$ on volume $L \ne L_0$ to $\beta_{L_0}$ on the reference volume $L_0$ by solving the  scaling equation \ref{eq:scaling2} 
\begin{equation}
    (L/a)^{1/\nu} |\beta_b /\beta^*_b -1| =  (L_0/a)^{1/\nu} |\beta_{L_0} /\beta^*_b -1|, 
\end{equation}
for second order scaling, and similarly for mFP scaling. We determine the infinite volume critical coupling $\beta_b^*$ and exponent $\nu$ or $\zeta$ by minimizing the deviation of $L\, M_{PS}(\beta_{L_0}(\beta_b);L)$ from the reference curve. We refine the reference curve by including other volumes and repeat the process. Finally, we test the stability of the FSS curve collapse by systematically removing data points that are farthest from the critical coupling. Since $L\,M_{PS}$ increases monotonically for $\beta_b <\beta_b^*$, this is easiest done by restricting the maximum value of $L\,M_{PS}$ included in the FSS fit.  

\Fig{fig:FSS} shows the result of the curve collapse analysis based on mFP scaling, while  
in the Supplemental Materials section shows it for second order scaling. Both fits are acceptable with small $\chi^2$. However, removing data furthest from criticality introduces a systematic change in the predicted exponents. In particular, the exponent $\nu$ of the second order scaling form increases steadily as the largest 10 of the available $L\,M_{PS}$ values are removed, as is shown on the right panel of Fig.~(2) 
in the Supplemental Materials section. The 
 predicted $\zeta$ values of the mFP fits also drift, but to a lesser extent.   

In any case, the exponent predicted by the second order scaling form is $\nu=0.61(5)$ or larger, significantly different from the first order discontinuity exponent $\nu_{\text{disc}}=1/d=0.25$. Thus, we can conclude that the phase transition that separates the conformal and SMG phases is continuous, which allows for
a new non-QCD like continuum theory to be defined from the SMG phase.
Taking into account both the FSS analysis from the strong coupling side and the spectrum from the weak coupling regime, we conclude that this system is most likely described by the mFP
scenario with an RG $\beta$ function that just touches zero. This then suggests that the SU(2) theory with $N_f=4$ flavors lies at the opening of the conformal window.

\subsection{Chiral symmetry}

\begin{figure*}[tbh]
    \centering
    \includegraphics[width=0.495\textwidth]{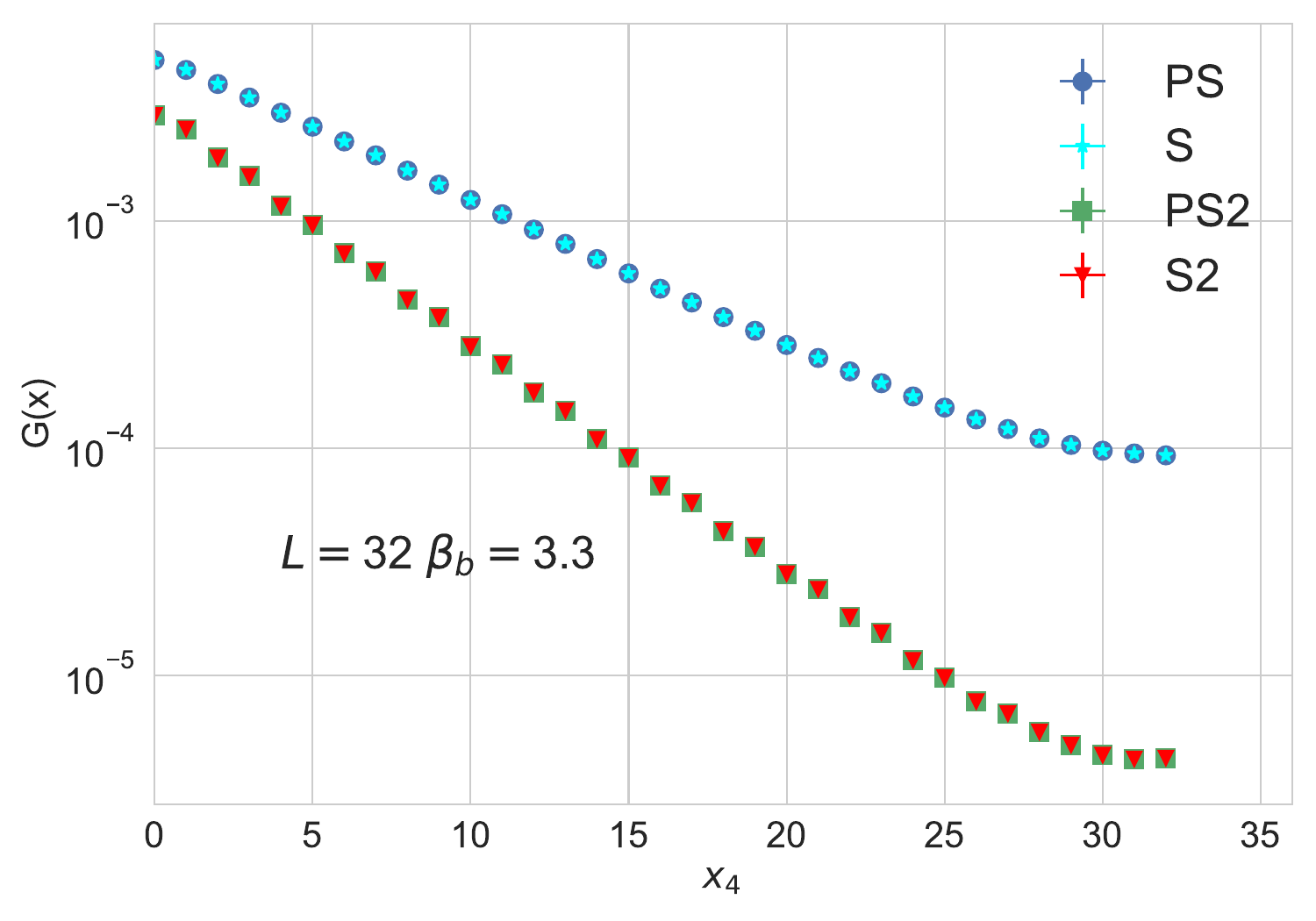}
    \includegraphics[width=0.495\textwidth]{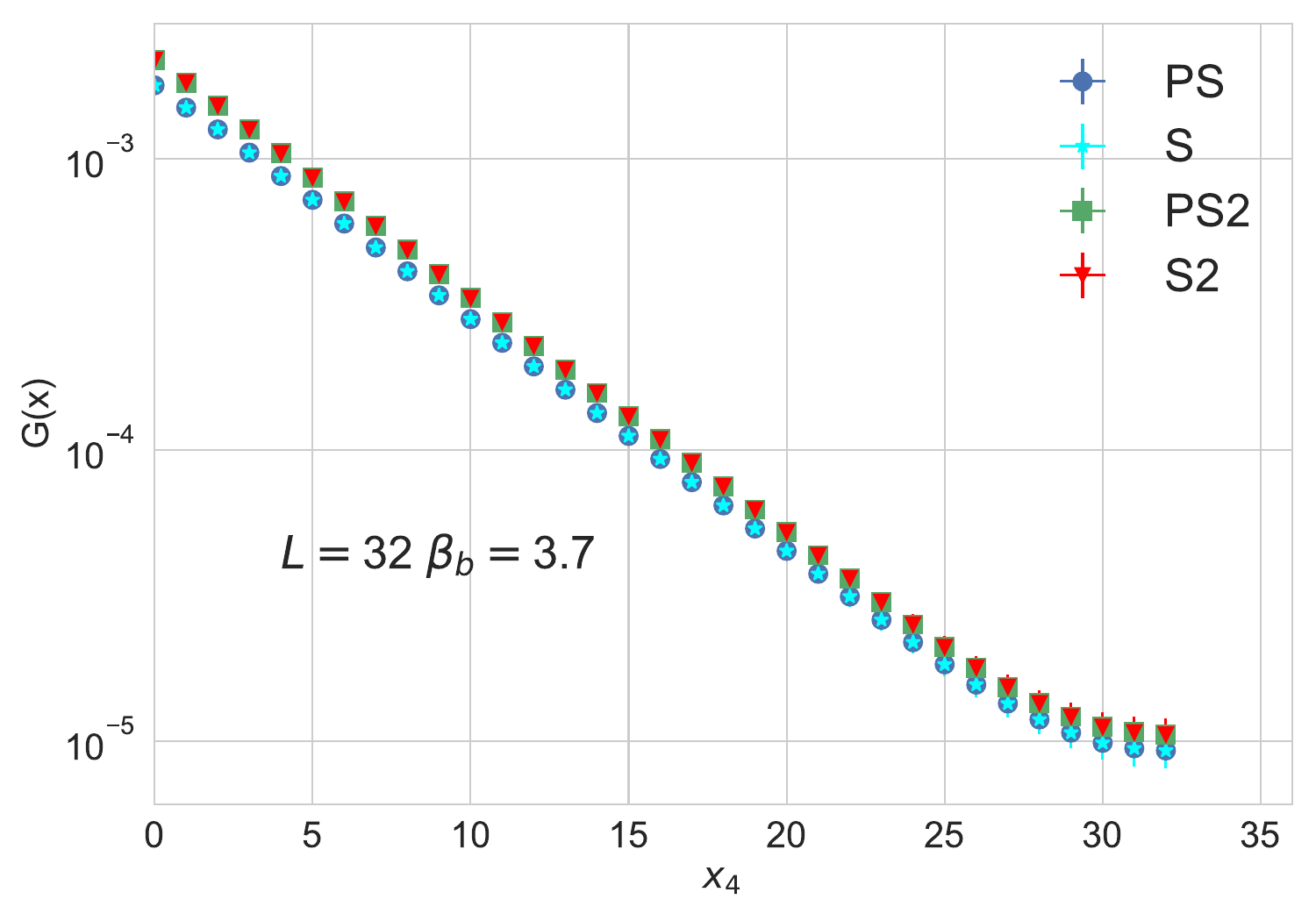}\\
    \caption{ The correlators of the two pseudoscalar operators PS and PS2, together with their scalar parity partners S and S2  on $L/a=32$ volumes. Left panel: $\beta_b=3.3$ in the SMG phase; right panel: $\beta_b=3.7$ in the conformal phase. The parity symmetry is unbroken, configuration-by-configuration, in both phases. The vector and axial correlators are similarly degenerate.}
    \label{fig:ChSym}
\end{figure*}

In this section we focus on possible chiral symmetry breaking in the theory. Clearly chiral
symmetry does not break in the conformal phase - as evidenced by the observed conformal hyperscaling
of the pseudoscalar (PS) and vector (V) states.~\footnote{There is no indication of any Goldstone boson that would signal spontaneous symmetry breaking. The ratio of $M_V/M_{PS}\approx 2.5$ but it does not diverge - the PS state is {\it not} a Goldstone boson.} What is more surprising is that we see no evidence for chiral symmetry breaking
at strong coupling in the SMG phase.

To quantify this claim, in \Fig{fig:ChSym} we show the correlators of the two pseudoscalar states (PS and PS2) and their parity partners (S and S2) in the strong coupling regime ($\beta_b=3.30$, left panel) and in the weak coupling phase ($\beta_b=3.70$, right panel). In both regimes, there is excellent parity degeneracy that is present configuration by configuration.  We observe the same degeneracy between the vector and axial vector states (see Fig.~(3) 
in the Supplementary Materials). Parity degeneracy, combined with the gapped spectrum, justifies our claim that the strong coupling phase while
gapped is not chirally broken -- it is an SMG phase.

\section{Discussion}
We have shown evidence that the theory of a single staggered fermion in the fundamental representation
of an SU(2) gauge group exhibits a two phase structure with a conformal phase at weak coupling
separated by a continuous phase transition from a gapped, confining and chirally
symmetric phase at strong coupling - an SMG phase. In fact our
results favor a situation where the RG $\beta$ function vanishes quadratically close to the new
fixed point consistent with the merging fixed point scenario discussed in \cite{Kaplan:2010zz,Vecchi:2010jz,Gorbenko:2018ncu} and associated with the lower boundary of
the conformal window. 

This suggests that the theory occupies a special point in theory
space. Indeed, this is the case --  theories of massless 
staggered fermions are invariant under a $Z_4$
symmetry which possesses a mod 2 't Hooft anomaly which is only cancelled for
multiples of two staggered fields or eight Dirac fields in the continuum limit.
The SU(2) theory discussed in this paper has the
minimal number of fermions in the UV that satisfy this constraint.

The infrared properties of the SU(2) gauge theory coupled to fermions in the
fundamental representation have been studied earlier using Wilson fermions \cite{Karavirta:2011zg,Amato:2015dqp}.  
In these works the authors see evidence of non-QCD behavior in the $N_f=4$ system but do not definitively identify either an IR conformal phase, or a chirally broken one. Instead, they try to interpret the results in terms of finite volume effects. To access larger renormalized couplings they are then forced to use strong bare coupling and hence encounter cut-off effects and a bulk transition. These results differ
from our work which uses staggered fermions.
It is a very interesting question if an SMG phase should (could) be universal between different fermion formulations. Using Wilson fermions to search for an SMG phase is
problematic since they do not retain the crucial $Z_4$ symmetry needed for the 't Hooft anomaly analysis. Indeed, Wilson fermions break the symmetry to $Z_2$ which automatically favors a bilinear, chiral
symmetry breaking condensate. However, we see no obvious obstruction to realizing an SMG phase
using overlap or domain wall fermions, provided they are equipped with (small) four fermion terms to guarantee
the required $Z_4$ symmetry. It would be very interesting to check this conclusion.
Certainly continuum theories of Weyl fermions exhibit a spin-$Z_4$ anomaly whose cancellation
picks out the same number of fermions as the staggered fermions we use.
 
In the future we would like to extend our
simulations to larger volumes to cement our confidence in the mFP scenario and verify the existence of a four-fermion condensate in the strong coupling phase.

\section{Acknowledgements}

Computations for this work were carried out in part on facilities of the USQCD Collaboration, which are funded by the Office of Science of the U.S. Department of Energy and  the RMACC Alpine supercomputer \cite{6866038}, which is supported by the National Science Foundation (awards No. ACI-1532235 and No. ACI-1532236), the University of Colorado Boulder, and Colorado State University. 

The numerical simulations were performed using the \texttt{Quantum EXpressions} (\texttt{QEX}) code \cite{Osborn:2017aci,Jin:2016ioq}\footnote{The \textsc{QEX} code is a lattice field theory framework written by James Osborn and Xiaoyong Jin in a general-purpose, multi-paradigm systems programming language called \textsc{Nim}. The open-source \textsc{QEX} code can be found at \url{https://github.com/jcosborn/qex}. The Pauli-Villars improvement was implemented by C.T. Peterson and can be found at \url{https://github.com/ctpeterson/qex}.}. We thank James Osborn and Xiaoyong Jin for their assistance with the code, and Curtis T. Peterson for getting us set up with \textsc{QEX}.

Anna Hasenfratz acknowledges support from DOE grant DE-SC0010005 and Simon Catterall from
DOE grant DE-SC0009998. AH thanks Oliver Witzel, Ethan Neil, and members of the LSD Collaboration for helpful discussions. SMC thanks Jay Hubisz for useful conversations. We are grateful to Cenke Xu who suggested that the $N_f=4$ SU(2) gauge system could exhibit a continuum SMG phase.

\bibliographystyle{unsrt}
\bibliography{bibliograph}

\cleardoublepage
\newpage

\onecolumngrid
\section*{Supplementary Materials}

\begin{table*}[tbh]
\centering
\begin{tabular}{| c | c | c | c |}
\hline
 state & spin$\otimes$taste & operator & meson  \\
  \hline
PS & $\gamma_5 \otimes \gamma_5$ 
   & $\sum_{\bar x} \chib(x)\chi(x) \eta_4(x)$& $\pi$ \\
S & $\gamma_0 \gamma_5 \otimes \gamma_0 \gamma_5$ 
   &$ \sum_{\bar x} \chib(x)\chi(x)$ & $a_0 \pm \pi$ \\
PS2 & $ \gamma_5 \otimes \gamma_k \gamma_5$ 
   & $\sum_{\bar x} \chib(x)U_k(x)\chi(x+k) \epsilon(x)\zeta_k(x)$ 
   & $\pi \pm a_0$\\
S2 &  $\gamma_0 \gamma_5 \otimes \gamma_i \gamma_j \,$ 
   &$\, \sum_{\bar x} \chib(x)U_k(x)\chi(x+k) \eta_k(x)\zeta_k(x)\epsilon(x) \,\,$ & $\,a_0 \pm \pi \,\,$ \\
V & $\gamma_k \otimes  \gamma_k$ 
   & $\sum_{\bar x} \chib(x)\chi(x) \eta_k(x)\zeta_k(x)\epsilon(x) $ & $\rho \pm b_1$ \\
A & $\gamma_0\gamma_k \otimes \gamma_0 \gamma_k$ 
   & $\sum_{\bar x} \chib(x)\chi(x) \eta_4(x)\zeta_4(x)\epsilon(x)\eta_k(x)$ & $a_1 \pm \rho$\\
\hline
\end{tabular}
\caption{Meson operators considered in this work. The first column is the notation used in the text, the second is the standard spin$\otimes$taste description and the third column is the representation in terms of staggered fields. The phase factors are $\eta_\mu(x) = (-1)^{\sum_{i=1}^{\mu-1}x_i}$, $\epsilon(x)=(-1)^{\sum_{i=1}^{D}x_i}$, and $\zeta_\mu(x)=(-1)^{\sum_{i=\mu+1}^{D}x_i}$.  The last column lists the mesons that the operators couple to in QCD. In QCD 
only the PS state creates a true Goldstone boson, the other $\pi$ states are lifted by taste breaking. \cite{Gupta:1990mr,Lee:1999zxa} }\label{tab:notation}
\end{table*}
\begin{figure}[tbh]
    \centering
    \includegraphics[width=0.495\textwidth]{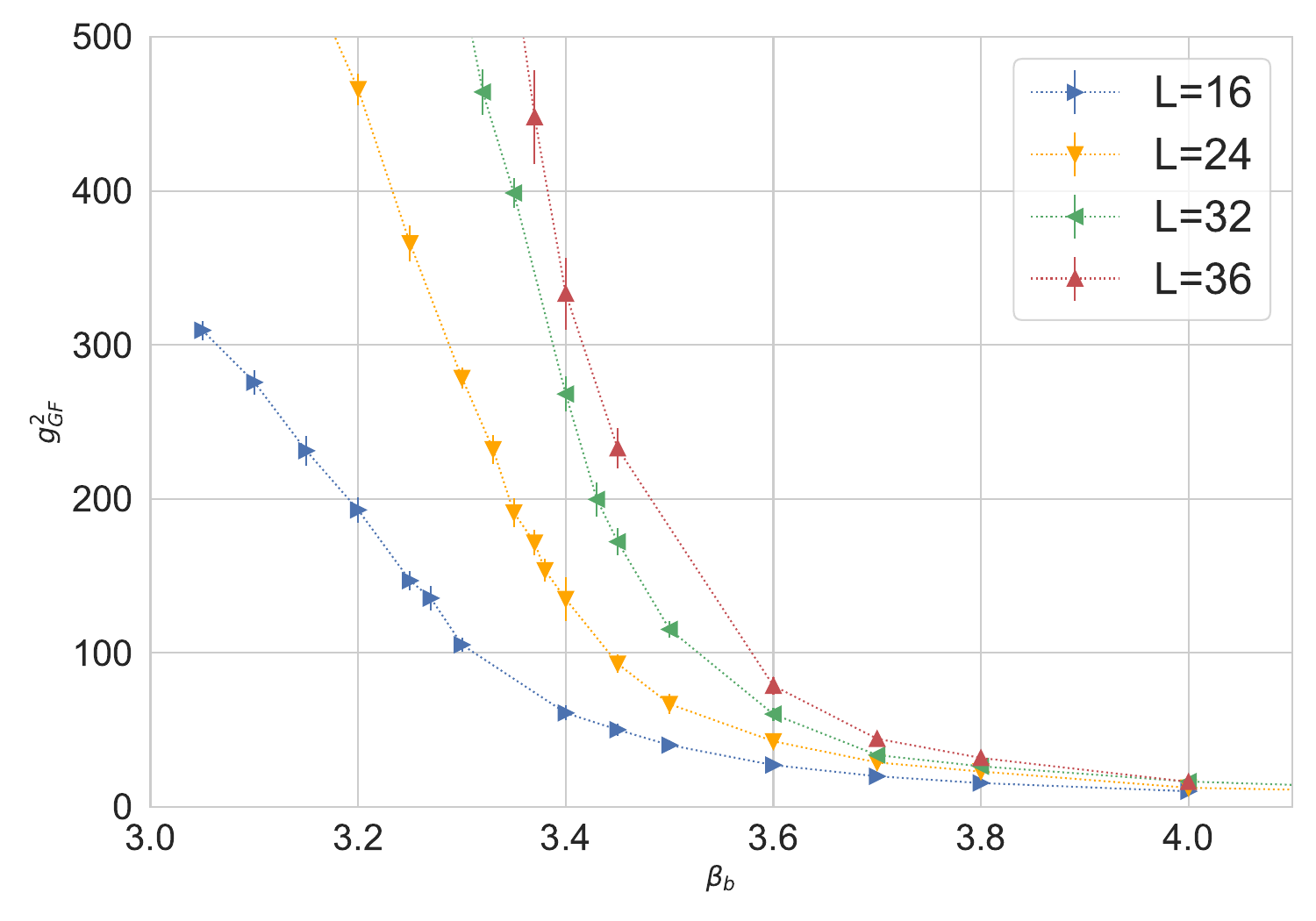}
    \caption{The finite volume gradient flow renormalized coupling $g^2_{GF}$ at flow time $c=\sqrt{8t}/L=0.5$ \cite{Luscher:2010iy}. Since $g^2_{GF}$ is RG invariant, its volume dependence at constant value indicates the direction of the RG flow in bare parameter space. As the plot shows the flow is always to the left (large volume to small), even though the system undergoes a phase transition around $\beta_b^* \approx 3.5$.\\
    The finite volume gradient flow renormalized coupling $g^2_{GF}$ has been used extensively to determine the RG $\beta$ function in asymptotically free systems. For details we refer the reader to some of the original literature \cite{Fodor:2012td,Fodor:2014cpa,Hasenfratz:2022yws}.} 
    \label{fig:g2_raw}
\end{figure}

\begin{figure*}[tbh]
    \centering
    \includegraphics[width=0.495\textwidth]{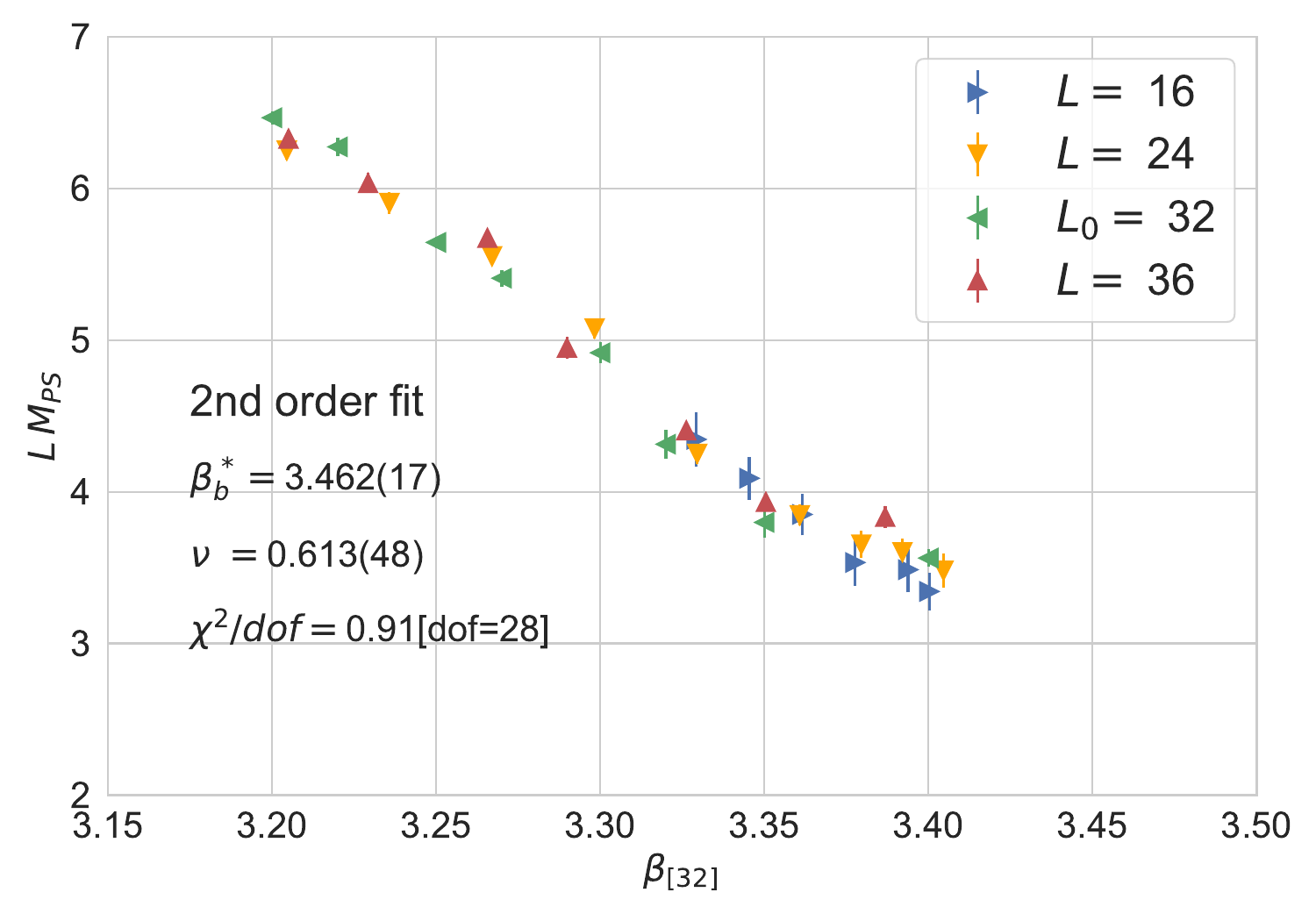}
    \includegraphics[width=0.495\textwidth]{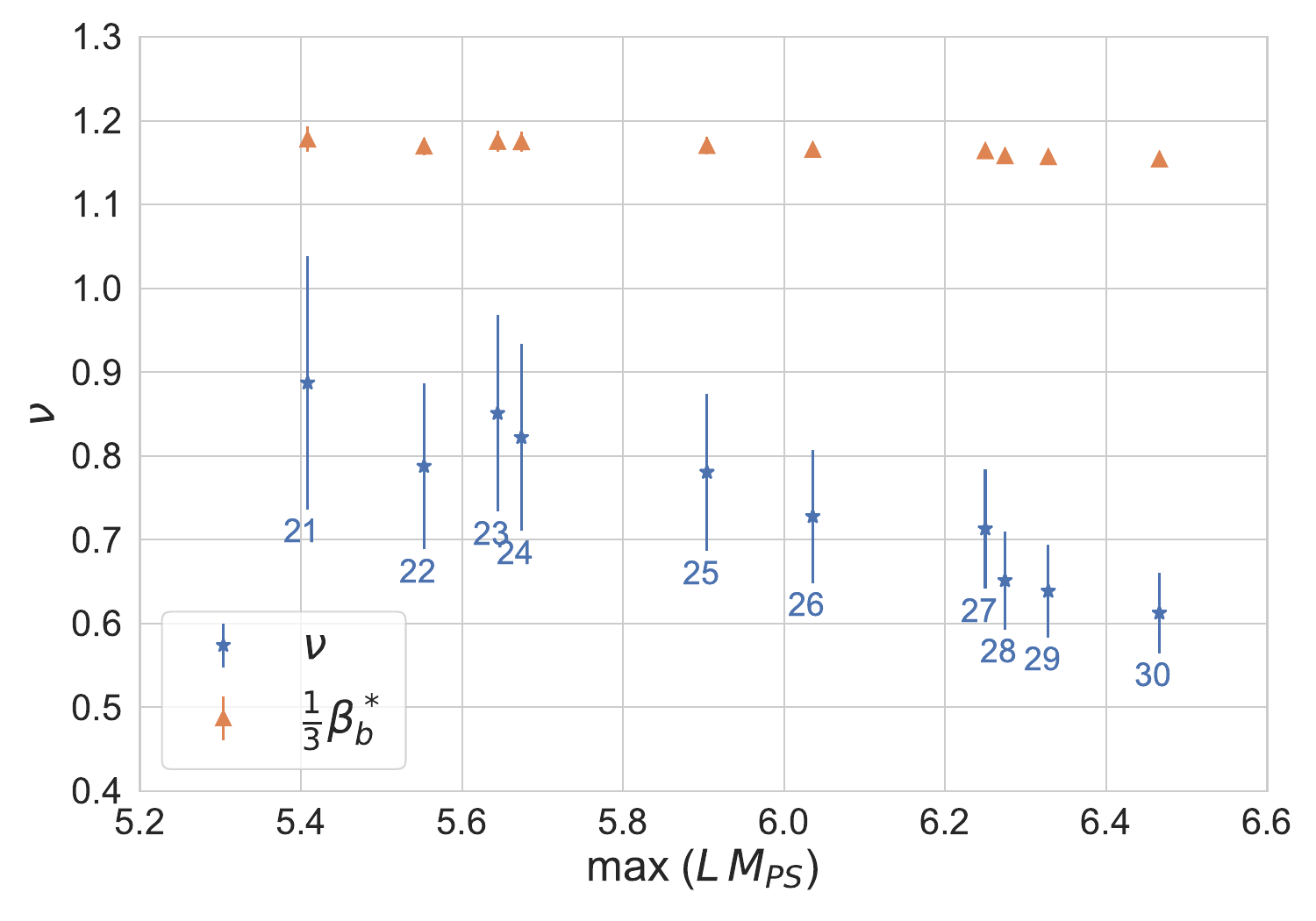}\\
    \caption{Finite size scaling plots using the operator $L\, M_{PS}$. The left panel shows the curve collapse fit assuming second order scaling. We present the results in terms of the coupling $\beta_b$ of the $L=32$ volume according to Eq. \ref{eq:scaling2}.  
    The right panel shows the sensitivity of the predicted $\nu$ exponent and critical coupling $\beta_b^*$ (divided by $3$ to better fit the plot) as the fit range is restricted using data closer and closer to the critical regime. The horizontal axis on the right panel corresponds to the maximum value of $L\,M_{PS}$ included in the fit, while the blue numbers under each data point denote the number of data points in the fit. The exponent $\nu$ increases as the fit range is reduced. The change is several standard deviation, suggesting that near the critical point the RG $\beta$ function    $\beta(g^2) \sim g^2/\nu$ is becoming flatter. } 
    \label{fig:2nd-orderb}
\end{figure*}

\begin{figure*}[tbh]
    \centering
    \includegraphics[width=0.495\textwidth]{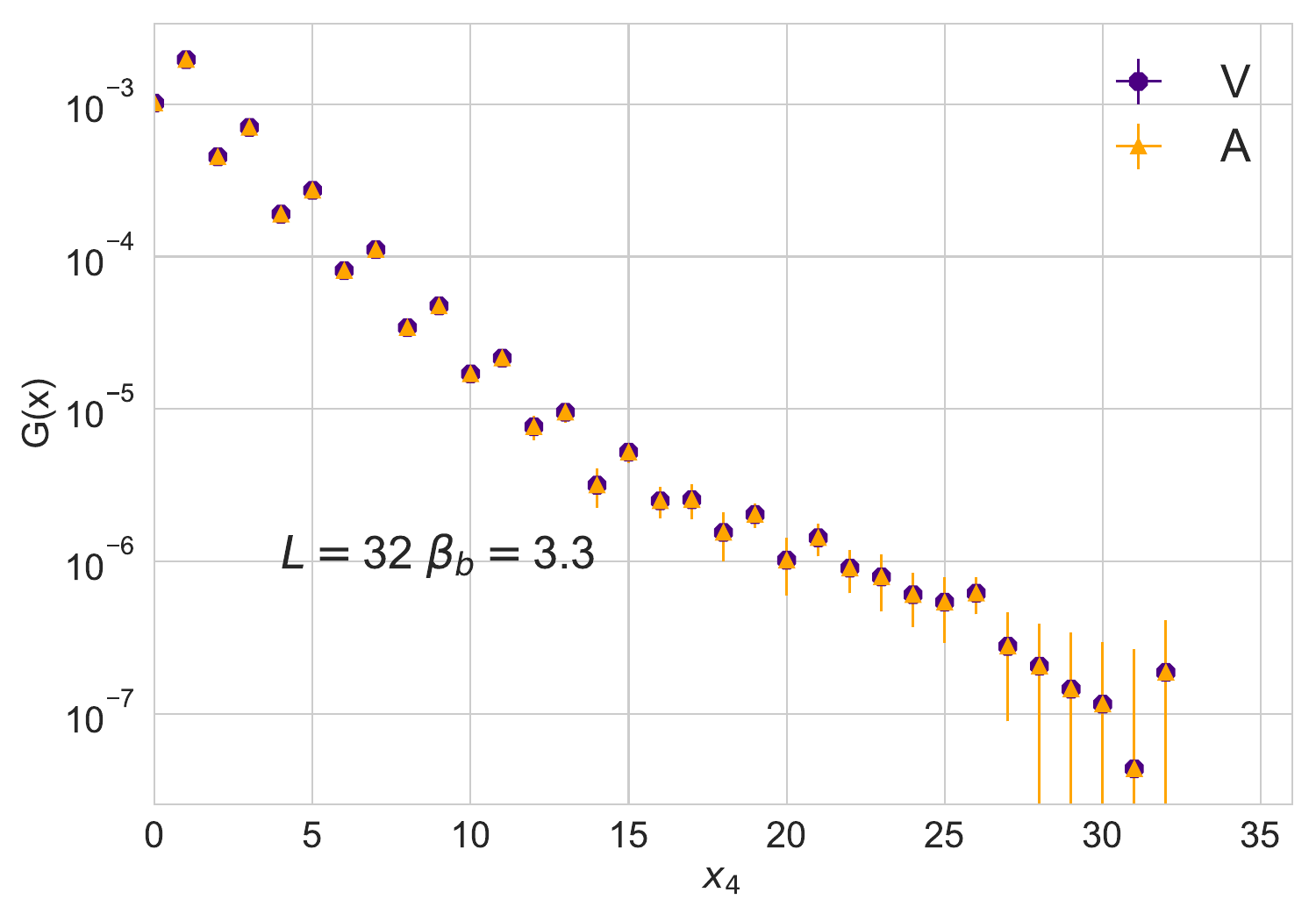}
    \includegraphics[width=0.495\textwidth]{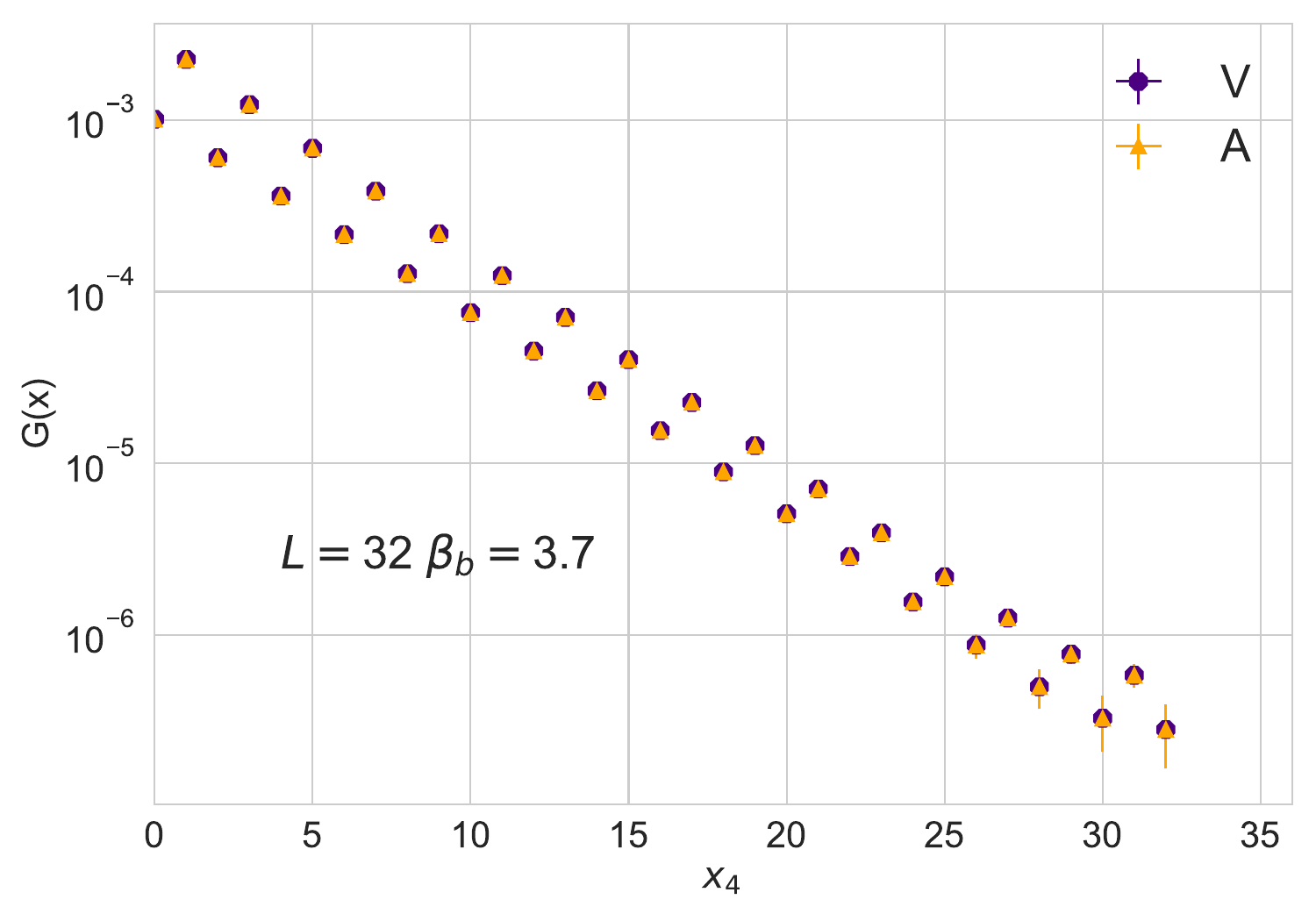}\\
    \caption{ The correlators of the vector and parity partner axial vector operators V and A on $L/a=32$ volumes. Left panel: $\beta_b=3.30$ in the weak coupling SMG phase; right panel: $\beta_b=3.70$ in the strong coupling conformal phase. The parity symmetry is unbroken, configuration-by-configuration, in both phases, explaining the perfect degeneracy even when the statistical errors are large.}
    \label{fig:ChSym_V}
\end{figure*}

\end{document}